\documentclass[12pt]{article}
\usepackage{amsmath}
\usepackage{amssymb}
\usepackage[mathscr]{eucal}
\usepackage[usenames]{color}
\usepackage{graphicx}
\usepackage{epstopdf}
\usepackage{subfigure}
\usepackage{graphics}
\usepackage{url}
\usepackage{enumitem}

\usepackage{url}
\usepackage[
bookmarks=true,
backref=true,
colorlinks=true,
linkcolor=blue,
citecolor=red,
urlcolor=green]{hyperref}

\newtheorem{proposition}{Proposition}
\newtheorem{theorem}{Theorem}
\newtheorem{corollary}{Corollary}
\newtheorem{remark}{Remark}

\numberwithin{equation}{section}
\numberwithin{theorem}{section}
\numberwithin{lemma}{section}
\numberwithin{proposition}{section}
\numberwithin{corollary}{section}
\numberwithin{remark}{section}
\numberwithin{definition}{section}

\newcommand{\semi}{\supset \hskip -5mm +}

\definecolor{darkolivegreen}{rgb}{0.333333, 0.419608, 0.1843140}

\newcommand{\dx}{\partial_x}
\newcommand{\dy}{\partial_y}
\newcommand{\dz}{\partial_z}
\newcommand{\dt}{\partial_t}
\newcommand{\du}{\partial_u}
\newcommand{\dv}{\partial_v}
\newcommand{\dw}{\partial_w}

\begin{document}

\title{\Large A Modified Davey--Stewartson System of Nonlinear Dust Acoustic Waves in (3+1)-Dimensions: Lie Symmetries and Exact Solutions
}

\author{
S. Gonul, Y. Hasanoglu, A. Tiryakioglu, Y. Calis,  C. Ozemir\thanks{Corresponding author, e-mail: ozemir@itu.edu.tr}\\
\small Department of Mathematics, Faculty of Science and Letters,\\
\small Istanbul Technical University, 34469 Istanbul,
Turkiye }

%%%%\date{24 February 2019}

\maketitle

\begin{abstract}
This article is devoted to the analysis of a modified Davey--Stewartson  system in three space dimensions, which was obtained in plasma physics for propagation of nonlinear dust acoustic waves. The system differs from the Davey--Stewartson systems available in the literature by an additional term which can be viewed as a constant complex  potential. We show that, under a certain condition on the parameters of the system, this term can be removed by a transformation. This restriction also separates the different realizations of   Lie symmetry algebra of the modified Davey--Stewartson system, which is identified as  semi-direct sum of a finite-dimensional algebra with a  Kac--Moody algebra. Having shed light on the group-theoretical properties of the system,  we present several results on the exact solutions of  generalized traveling wave type, some of which are line solitons and kink solitons on  planes in space. We finalize by analysing the stability of traveling wave solutions.  
\end{abstract}

\section{Introduction}
The aim of this paper is to investigate the system 
\begin{subequations}\label{wang}
\begin{eqnarray}     &&i\Phi_{\tau}+\alpha_1\Phi_{\xi\xi}+\alpha_2(\Phi_{\eta\eta}+\Phi_{\zeta\zeta})+\alpha_3|\Phi|^2\Phi+\alpha_4 \Phi \varphi+i\alpha_5\Phi=0, \\
&&\varphi_{\xi\xi}-\beta_2(\varphi_{\eta\eta}+\varphi_{\zeta\zeta})=\beta_1\Big((|\Phi|^2)_{\eta\eta}+(|\Phi|^2)_{\zeta\zeta}\Big)
\end{eqnarray}
\end{subequations}
which was derived in \cite{wang2009modulational} via asymptotic analysis of  nonlinear dust acoustic  waves in dusty plasma
consisting of Boltzmann-distributed electrons, ions and positively charged dust grain. Here $\alpha_i$, $i=1,..,5$ and $\beta_1$, $\beta_2$ are real constants.   $\Phi$ and $\varphi$ are short and long wave components of the propagation, respectively.  Due the term $i\alpha_5 \Phi$, the authors call \eqref{wang} the  three-dimensional
modified Davey–Stewartson (3D-MDS) equations.  In the current article, we focus on group-theoretical properties of this system and explore its infinite-dimensional Lie symmetry algebra.  Furthermore, we present results on exact plane wave solutions of the equation.

Similar to this system, the work \cite{ozemir2020davey} is devoted to the Lie symmetries of a Davey--Stewartson system in $(3+1)$ dimensions in the form 
\begin{subequations}\label{31DS}
\begin{eqnarray}
   \label{31DSa}  &&i\psi_t+\psi_{xx}+a_1\psi_{yy}+\psi_{zz}=a_2|\psi|^2\psi+\psi w, \\
   \label{31DSb}  &&w_{xx}+b_1w_{yy}+w_{zz}=b_2(|\psi|^2)_{yy}.
\end{eqnarray}
\end{subequations}
In order to be able to compare the results we shall obtain with our analysis for \eqref{wang} to the results available in \cite{ozemir2020davey} for \eqref{31DS}, 
we first transform \eqref{wang} via
\begin{equation}
\tau=\frac{t}{\alpha_2}, \quad \xi=y, \quad \eta=x, \quad \zeta=z, \quad \Phi=\psi, \quad \varphi=-\frac{\alpha_2}{\alpha_4}w.
\end{equation}
We get 
\begin{subequations}\label{31MDS}
\begin{eqnarray}
   \label{31MDSa}  &&i\psi_t+\psi_{xx}+a_1\psi_{yy}+\psi_{zz}=a_2|\psi|^2\psi+\psi w+ia_3 \psi, \\
   \label{31MDSb}  &&w_{xx}+b_1w_{yy}+w_{zz}=b_2\big [ (|\psi|^2)_{xx} +(|\psi|^2)_{zz}\big]
\end{eqnarray}
\end{subequations}
with 
\begin{equation}
a_1=\frac{\alpha_1}{\alpha_2}, \quad a_2=-\frac{\alpha_3}{\alpha_2}, \quad a_3=-\frac{\alpha_5}{\alpha_2}, \quad b_1=-\frac{1}{\beta_2}, \quad b_2=\frac{\beta_1\alpha_4}{\beta_2\alpha_2}.
\end{equation}
Let us note that in our analysis we assumed  $\alpha_1$, $\alpha_2$, $\alpha_3$, $\alpha_4$, $\beta_1$ and $\beta_2$ to be nonzero, but did not put this condition on  $\alpha_5$. Therefore in Eq. \eqref{31MDS}  $a_1$, $a_2$, $b_1$, $b_2$ are nonzero real constants and $a_3$ is a real constant which can assume the value zero.

In the literature of PDEs of mathematical physics, Davey--Stewartson (DS) system encompasses a significant volume of works from many perspectives.  It first appeared in  \cite{daveystewartson} in two space dimensions by a perturbation expansion, as governing equations for the evolution of a wave packet in  water of finite depth. Let us note that it is a  special case of the Benney--Roskes system derived in \cite{benneyroskes}. Ref.  \cite{fokas1983inverse} studies integrability of the Davey--Stewartson system through inverse scattering transform. On the other hand, well-posedness of the initial value problem was studied in \cite{ghidaglia1990initial} and was continued by \cite{linares1993davey}.

Champagne and Winternitz  studied the Lie symmetry algebra of the DS system in $(2+1)$ dimensions
\begin{subequations}\label{2DDS}
\begin{eqnarray}
   \label{2DDSa}  &&i\psi_t+\psi_{xx}+\epsilon_1\psi_{yy}=\epsilon_2|\psi|^2\psi+\psi w, \\
   \label{2DDSb}  &&w_{xx}+\delta_1w_{yy}=\delta_2(|\psi|^2)_{yy}
\end{eqnarray}
\end{subequations}
in \cite{champagnewinternitz1988} and they showed that it has an  infinite-dimensional Lie symmetry algebra of Kac--Moody--Virasoro (KMV) type exactly in the integrable case; i.e., when $\delta_1=-\epsilon_1=\mp 1$. The DS system \eqref{2DDS} has two components of vibrations perpendicular to each other, the short wave $\psi$ and the long wave $w$. In the context of solid mechanics, in $(2+1)$ dimensions the authors of \cite{babaoglu2004two} derive a system which they call "generalized Davey--
Stewartson equations" (GDS) that include three wave components as short--long--long wave interactions. The symmetry algebra and exact solutions of the GDS system are studied in \cite{gungor2006generalized} and \cite{li2008symmetry}. Some of the integrable systems in $(2+1)$  dimensions known to possess Lie algebras of KMV type. This motivated the search for this type of invariance algebras in $(2+1)$ dimensions as a glimpse  of integrability. For results in this direction, we can cite  \cite{gungor2006virasoro} and other related references in \cite{ozemir2020davey}. For the works  on the derivation of DS equations in $(2+1)$ dimensions in different contexts, we would like to refer to the references \cite{nishinari1993new,nishinari1994new,lin2001new,panguetna2017two,duan2004two}.  

Regarding the works on physical models in which $(3+1)$-dimensional  Davey--Stewartson system appears as the governing equations,  first we can mention Ref.  \cite{carbonaro2012three} in which the author derives a  $(3+1)$-dimensional DS system  for  an electron-acoustic wave in a collisionless unmagnetized plasma. Another three dimensional Davey–Stewartson equation in
unmagnetized dusty plasmas is obtained in \cite{xue2004modulational}. The work \cite{leblond1999electromagnetic} reported that  nonlinear modulation of an electromagnetic localized pulse in a
saturated bulk ferromagnetic medium is governed by a  three-dimensional DS  system. Last we can mention Ref. \cite{tabi2018electronegative}, in which a three-dimensional DS system appears as the interaction equation for the ion-acoustic waves. 

We see that the 3D-MDS system \eqref{wang} has not been considered in the literature from a group-theoretical point of view and for  analytical solutions. In Section 2, considering the system in the form \eqref{31MDS}, we first investigate  transformability of \eqref{31MDS} to the same equation  with $a_3=0$. After that,  we obtain the infinite-dimensional Lie algebra structure of \eqref{31MDS} and exhibit the correspondence with the Lie algebra of the system \eqref{31DS} in a special case of the constants. The Lie algebra appears as the semi-direct sum of a Kac--Moody algebra with a finite-dimensional Lie algebra.  In Section 3 we provide reductions of \eqref{31MDS} to ODEs that determine generalized traveling wave solutions. One of the reduced equations we obtain was previously studied  extensively for a different PDE in the literature, therefore those results serve also for the current problem. Specifically, we obtain some periodic solutions and solutions in the forms of line solitons and kink solitons.  We finalize the manuscript with  some results on stability of these exact solutions.

\section{Group-Theoretical Analysis}
\subsection{Transformation}
Comparing the forms of \eqref{31MDS} and \eqref{31DS}, we first check whether by a transformation we can set the coefficient $a_3$ in \eqref{31MDSa} to zero. To achieve this, we perform a transformation of the independent and dependent variables as  $(x,y,z,t,\psi,w)\rightarrow(X,Y,Z,T,\Psi,W)$, in the form 
\begin{equation}\label{form1}
X=X(x,y,z,t), \quad Y=Y(x,y,z,t), \quad Z=Z(x,y,z,t), \quad T=T(t),
\end{equation}
with $\dot T = dT/dt  >0 $ to preserve direction of time and 
\begin{equation}\label{form2}
\begin{split}
\psi(x,y,z,t)&=R(x,y,z,t)e^{i\theta(x,y,z,t)}\Psi(X,Y,Z,T), \\
w(x,y,z,t)&=P(x,y,z,t) W(X,Y,Z,T)+S(x,y,z,t).
\end{split}
\end{equation}
Her $R$, $\theta$, $P$, $S$ and $W$ are real-valued functions and $\Psi$ is complex-valued. We plug \eqref{form1} and \eqref{form2} in \eqref{31MDS}
aiming at obtaining
\begin{subequations}\label{31MDS2}
\begin{eqnarray}
   \label{31MDS2a}  &&i\Psi_T+\Psi_{XX}+a_1\Psi_{YY}+\Psi_{ZZ}=a_2|\Psi|^2\Psi+\Psi W, \\
   \label{31MDS2b}  &&W_{XX}+b_1W_{YY}+W_{ZZ}=b_2\big [ (|\Psi|^2)_{XX} +(|\Psi|^2)_{ZZ}\big].
\end{eqnarray}
\end{subequations}

Considering coefficients of $\Psi_T$, $\Psi_{XX}$, $\Psi_{YY}$, $\Psi_{ZZ}$ we see that we must have
\begin{subequations}\label{transxyz}
\begin{eqnarray}
   \label{transx}  && X=\epsilon_1 \sqrt{\dot T(t)} \,   x+\mu(t),\\
   \label{transy}  && Y=\epsilon_2 \sqrt{\dot T(t)} \,   y+\nu(t),\\
   \label{transz}  && Z=\epsilon_3 \sqrt{\dot T(t)} \,   z+\delta(t)
\end{eqnarray}
\end{subequations}
and the terms $|\Psi|^2\Psi$ and $\Psi  W$ require
\begin{equation}
R=\epsilon_4 \sqrt{\dot T (t)}, \qquad P=\dot T (t)
\end{equation}
with $\epsilon_1=\mp 1$, $\epsilon_2=\mp 1$, $\epsilon_3=\mp 1$, $\epsilon_4=\mp 1$. The additional terms appearing as coefficients of $\Psi$, $\Psi_X$, $\Psi_Y$, $\Psi_Z$ determine the remaining set of restrictions on the transformation functions as 
\begin{subequations}\label{trans}
\begin{eqnarray}
\label{transa}  &&  \theta_x+\frac{\ddot T }{4\dot T }x+\frac{\dot \mu}{2\epsilon_1 \dot T ^{1/2}}   =0,  \\
   \label{transb}  && \theta_y+\frac{\ddot T }{4a_1\dot T }y+\frac{\dot \nu}{2a_1\epsilon_2 \dot T  ^{1/2}}  =0,  \\
    \label{transc} &&  \theta_z+\frac{\ddot T }{4\dot T }z+\frac{\dot \delta}{2\epsilon_3 \dot T  ^{1/2}}  =0, \\
   \label{transd}  && \theta_{xx}+a_1 \theta_{yy}+\theta_{zz}+\frac{\ddot T}{2\dot T}-a_3=0,\\
   \label{transe}  && \theta_t+\theta_x^2+a_1\theta_y^2+\theta_z^2+S=0, \\
   && S_{xx}+b_1 S_{yy}+ S_{zz}=0. \label{transf} 
\end{eqnarray}
\end{subequations}
Using equations \eqref{transa}-\eqref{transd} we determine that 
\begin{equation}
T(t)=T_1 e^{-4a_3 t}+T_0, 
\end{equation}
with arbitrary constants $T_0$, $T_1$ and
\begin{equation}
\theta(x,y,z,t)=-\frac{\ddot T}{8 \dot T} \Big(x^2+\frac{y^2}{a_1}+z^2\Big)-\frac{1}{2\dot T^{1/2}} \Big(\frac{1}{\epsilon_1} \dot \mu x+\frac{1}{a_1\epsilon_2} \dot \nu y+\frac{1}{\epsilon_3} \dot \delta z \Big)+\sigma(t)
\end{equation}
where $\sigma(t)$ is an arbitrary function. As $\dot T =-4 a_3 T_1e^{-4a_3 t}$,  $T_1$ can be chosen properly depending on the sign of $a_3$ to satisfy the condition $\dot T >0$. The  restriction \eqref{transe} provides 
\begin{subequations}\label{munudeltatheta0}
\begin{eqnarray}
\label{mu}    \mu(t)&=&\mu_1 e^{-4a_3 t}+\mu_0,  \\
   \label{nu}   \nu(t)&=&\nu_1 e^{-4a_3 t}+\nu_0,  \\
    \label{delta}   \delta(t)&=&\delta_1 e^{-4a_3 t}+\delta_0,  \\
      S(x,y,z)&=&-a_3^2 (x^2+\frac{1}{a_1}y^2+z^2)
   \end{eqnarray}
\end{subequations}
and 
\begin{equation}
\sigma(t)=\frac{-1}{4T_1}\Big( \mu_1^2+\frac{\nu_1^2}{a_1}+\delta_1^2\Big) e^{-4 a_3 t}+\theta_0
\end{equation}
where $\theta_0$ is an arbitrary constant. Finally, from \eqref{transf} we obtain the last condition of transformability as 
\begin{equation}\label{cond}
b_1+2a_1=0,
\end{equation}
which corresponds to $2\alpha_1\beta_2=\alpha_2.$

\begin{proposition}
When $a_3\neq 0$, if $b_1+2a_1=0$, the transformation from \eqref{31MDS} to \eqref{31MDS2} is achieved through
\begin{equation}\label{form3}
\begin{split}
\psi(x,y,z,t)&=2\epsilon_4 \sqrt{- a_3 T_1}e^{-2a_3 t}\,\Psi(X,Y,Z,T) \exp\Bigg\{i\Bigg[\frac{a_3}{2}\Big(x^2+\frac{y^2}{a_1}+z^2\Big)\\
&+\sqrt{-\frac{a_3}{T_1}}\Big(\frac{\mu_1}{\epsilon_1}x+\frac{\nu_1}{a_1\epsilon_2}y+\frac{\delta_1}{\epsilon_3}z\Big)e^{-2a_3 t}-\frac{1}{4T_1}\Big( \mu_1^2+\frac{\nu_1^2}{a_1}+\delta_1^2\Big) e^{-4 a_3 t}+\theta_0\Bigg]\Bigg\},\\
w(x,y,z,t)&=-4a_3T_1e^{-4a_3 t} W(X,Y,Z,T)-a_3^2 (x^2+\frac{1}{a_1}y^2+z^2)
\end{split}
\end{equation}
where 
\begin{subequations}\label{transxyz2}
\begin{eqnarray}
   \label{transt2}  &&T=T_1 e^{-4a_3 t}+T_0, \\
   \label{transx2}  && X=2\epsilon_1 \sqrt{- a_3 T_1}e^{-2a_3 t} x+\mu_1 e^{-4a_3t} +\mu_0,\\
   \label{transy2}  && Y=2\epsilon_2 \sqrt{- a_3 T_1}e^{-2a_3 t} y+\nu_1 e^{-4a_3t} +\nu_0,\\
   \label{transz2}  && Z=2\epsilon_3 \sqrt{- a_3 T_1}e^{-2a_3 t} z+\delta_1 e^{-4a_3t} +\delta_0.
\end{eqnarray}
\end{subequations}
\end{proposition}
Let us note that there is an additional set of transformations with $x$ and $z$ replaced on the right hand sides of equations \eqref{form3} and \eqref{transxyz2}. 
\subsection{Lie symmetry algebra}
For $\psi=u+iv$ we express \eqref{31MDS} as the system
\begin{subequations}\label{Triple}
\begin{eqnarray}
   \label{triple1}  &&\Delta_1=u_t+v_{xx}+a_1v_{yy}+v_{zz}-a_2v(u^2+v^2)-vw-a_3 u =0, \\
   \label{triple2}  &&\Delta_2=-v_t+u_{xx}+a_1u_{yy}+u_{zz}-a_2u(u^2+v^2)-uw+a_3 v=0,\\
   \label{triple3}  &&\Delta_3=w_{xx}+b_1w_{yy}+w_{zz}-b_2(u^2+v^2)_{xx}-b_2(u^2+v^2)_{zz}=0.
\end{eqnarray}
\end{subequations}
Lie group of transformations that leave this system invariant are flows of the  vector fields of the form 
\begin{equation}
V=\sigma_1\dt+\sigma_2\dx+\sigma_3\dy+\sigma_4 \dz+\phi_1\du+\phi_2\dv+\phi_3\dw.
\end{equation}
The coefficients in $V$ are functions of the variables $(t,x,y,z,u,v,w)$ and $V$  is a symmetry generator if its second order prolongation annihilates  \eqref{Triple} on the solution surface:
\begin{equation}\label{inf-inv}
  \mathsf {Pr}^{(2)}V(\Delta_i)\big\vert_{\Delta_j=0}=0,\quad i,j=1,2,3.
\end{equation}

From the determining equations provided by the invariance condition \eqref{inf-inv}, we determine the coefficients of the vector field $V$. Before stating the main result on the symmetry algebra of the system \eqref{31MDS}, let us introduce the vector fields
\begin{subequations}
\begin{eqnarray}
\mathcal{X}_1&=&\dt,\\
\mathcal{X}_2&=&2t\dt+x\dx+y\dy+z\dz-u\du-v\dv-2w\dw,\\
\mathcal{\tilde X}_2 &=&e^{4a_3t}\Bigg\{\frac{1}{2a_3}\dt+x\dx+y\dy+z\dz-\Big[u+a_3\Big(x^2+\frac{y^2}{a_1}+z^2\Big)v\Big]\du  \nonumber\\
&+&\Big[-v+a_3\Big(x^2+\frac{y^2}{a_1}+z^2\Big)u\Big]\dv-\Big[2w+4a_3^2\Big(x^2+\frac{y^2}{a_1}+z^2\Big)\Big]\dw\Bigg\},\\
\mathcal{X}_3&=&z\dx-x\dz,
\end{eqnarray}
\end{subequations}
and also 
\begin{subequations}
\begin{eqnarray}
\mathcal{Y}(g)&=&g(t) \dx-\frac{x}{2}\big[\,g'(t)(v\du-u\dv)+g''(t)\dw\big],\\
\mathcal{Z}(h)&=&h(t) \dy-\frac{y}{2a_1}\big[\,h'(t)(v\du-u\dv)+h''(t)\dw\big],\\
\mathcal{Q}(k)&=&k(t) \dz-\frac{z}{2}\big[\,k'(t)(v\du-u\dv)+k''(t)\dw\big],\\
\mathcal{W}(m)&=&m(t)(v\du-u\dv)+m'(t) \,\dw
\end{eqnarray}
\end{subequations}
where $g(t)$, $h(t)$, $k(t)$ and $m(t)$ are arbitrary functions. 

The commutation relations are given below. The ones involving $\mathcal{X}_i$'s, $i=1,2,3$ are,
\begin{align}
&[\mathcal{X}_1,\mathcal{X}_2]=2\mathcal{X}_1,  &&[\mathcal{X}_1,\mathcal{Y}(g)]=\mathcal{Y}(g'), && [\mathcal{X}_1,\mathcal{Z}(h)]=\mathcal{Z}(h'), \nonumber\\
&[\mathcal{X}_1,\mathcal{Q}(k)]=\mathcal{Q}(k'),  & &[\mathcal{X}_1,\mathcal{W}(m)]=\mathcal{W}(m'),\nonumber\\
&[\mathcal{X}_2,\mathcal{Y}(g)]=\mathcal{Y}(2tg'-g),  &&[\mathcal{X}_2,\mathcal{Z}(h)]=\mathcal{Z}(2th'-h),\\
&[\mathcal{X}_2,\mathcal{Q}(k)]=\mathcal{Q}(2tk'-k),  &&[\mathcal{X}_2,\mathcal{W}(m)]=\mathcal{W}(2tm'),\nonumber\\
&[\mathcal{X}_3,\mathcal{Y}(g)]=\mathcal{Q}(g),  &&[\mathcal{X}_3,\mathcal{Q}(k)]=-\mathcal{Y}(k), \nonumber
\end{align}
and the remaining nonzero relations are,
\begin{align}
&[\mathcal{Y}(g_1),\mathcal{Y}(g_2)]=\mathcal{W}\big(\frac{1}{2}(g_1'g_2-g_1g_2')\big),\nonumber\\
&[\mathcal{Z}(h_1),\mathcal{Z}(h_2)]=\mathcal{W}\big(\frac{1}{2a_1}(h_1'h_2-h_1h_2')\big),\\
&[\mathcal{Q}(k_1),\mathcal{Q}(k_2)]=\mathcal{W}\big(\frac{1}{2}(k_1'k_2-k_1k_2')\big), \nonumber
\end{align}
and finally, just to state explicitly, the vanishing commutations are,
\begin{align}
&[\mathcal{X}_1,\mathcal{X}_3]=[\mathcal{X}_2,\mathcal{X}_3]=[\mathcal{X}_3,\mathcal{Z}(h)]=[\mathcal{X}_3,\mathcal{W}(m)]=0, \nonumber\\
&[\mathcal{Y}(g),\mathcal{Z}(h)]=[\mathcal{Y}(g),\mathcal{Q}(k)]=[\mathcal{Y}(g),\mathcal{W}(m)]=0,\\
&[\mathcal{Z}(h),\mathcal{Q}(k)]=[\mathcal{Z}(h),\mathcal{W}(m)]=[\mathcal{Q}(k),\mathcal{W}(m)]=[\mathcal{W}(m_1),\mathcal{W}(m_2)]=0. \nonumber
\end{align}
We separately present the relations including $\mathcal{\tilde X}_2$ as 
\begin{align}
&[\mathcal{X}_1,\mathcal{\tilde X}_2]=4 a_3 \mathcal{\tilde X}_2,  && [\mathcal{\tilde X}_2,\mathcal{X}_3]=0,   \nonumber \\
&[\mathcal{\tilde X}_2,\mathcal{Y}(g)]=\mathcal{Y}\Big(\frac{e^{4a_3t}}{2a_3}(g'-2a_3 g)\Big),  &&[\mathcal{\tilde X}_2,\mathcal{Z}(h)]=\mathcal{Z}\Big(\frac{e^{4a_3t}}{2a_3}(h'-2a_3 h)\Big),\\
&[\mathcal{\tilde X}_2,\mathcal{Q}(k)]=\mathcal{Q}\Big(\frac{e^{4a_3t}}{2a_3}(k'-2a_3 k)\Big),  &&[\mathcal{\tilde X}_2,\mathcal{W}(m)]=\mathcal{W}\Big(\frac{e^{4a_3t}}{2a_3}m'\Big).  \nonumber
\end{align}

Now, skipping the details of well-known  procedure of determining the Lie algebra of a differential equation which led to the symmetry generators above, we state the main result on the symmetry algebra of the system \eqref{31MDS} as follows. 

\begin{theorem}
\begin{itemize}
\item[(i)] For any values of the constants $a_1$, $a_2$, $a_3$, $b_1$, $b_2$, system \eqref{31MDS} admits the symmetry generators $\mathcal{Y}(g)$, $\mathcal{Z}(h)$, $\mathcal{Q}(k)$, $\mathcal{W}(m)$. Therefore, the symmetry algebra of the (3+1)-dimensional modified Davey--Stewartson system \eqref{31MDS} is infinite dimensional.
\item[(ii)] If $a_3=0$,  Lie algebra $L_1$ of \eqref{31MDS}  is
\begin{equation}
L_1=\{\{\mathcal{X}_1,\mathcal{X}_2\}\oplus \mathcal{X}_3\}\semi\{\mathcal{Y}(g), \mathcal{Z}(h), \mathcal{Q}(k), \mathcal{W}(m)\}
\end{equation}
where $\{\mathcal{X}_1,\mathcal{X}_2\}\oplus \mathcal{X}_3$ is a three-dimensional Lie algebra.
\item[(iii)] If $a_3\neq 0$ and $b_1=-2a_1$,  Lie algebra $L_2$ of \eqref{31MDS} is 
\begin{equation}
L_2=\{\{\mathcal{X}_1,\mathcal{\tilde X}_2\}\oplus \mathcal{X}_3\}\semi\{\mathcal{Y}(g), \mathcal{Z}(h), \mathcal{Q}(k), \mathcal{W}(m)\}
\end{equation}
where $\{\mathcal{X}_1,\mathcal{\tilde X}_2\}\oplus \mathcal{X}_3$ is a three-dimensional Lie algebra.
\item[(iv)] If $a_3\neq 0$ and $b_1\neq -2a_1$,  Lie algebra $L_3$ of \eqref{31MDS} is generated by
\begin{equation}
L_3=\{\mathcal{X}_1 \oplus \mathcal{X}_3\}\semi\{\mathcal{Y}(g), \mathcal{Z}(h), \mathcal{Q}(k), \mathcal{W}(m)\}
\end{equation}
where $\mathcal{X}_1\oplus \mathcal{X}_3$ is the two-dimensional abelian  Lie algebra.
\end{itemize}
\noindent In all of these cases, $\{\mathcal{Y}(g), \mathcal{Z}(h), \mathcal{Q}(k), \mathcal{W}(m)\}$ is a Kac--Moody algebra being a non-Abelian ideal.
\end{theorem}

Consider the three dimensional Lie algebras $S_1=\{\mathcal{X}_1,\mathcal{X}_2\}\oplus \mathcal{X}_3$ and $S_2=\{\mathcal{X}_1,\mathcal{\tilde X}_2\}\oplus \mathcal{X}_3$. If we make a change of basis in $S_2$ in the form $\mathcal{X}_1=-2a_3\mathcal{P}_2$, $\mathcal{\tilde X}_2=\mathcal{P}_1$, $\mathcal{X}_3=\mathcal{P}_3$, $S_1$ and $S_2$ satisfy the same commutation relations, therefore they are isomorphic. Further, the Lie symmetry algebra of the system \eqref{31DS} is investigated in \cite{ozemir2020davey} and reported to be exactly  of the structure $\{\{\mathcal{X}_1,\mathcal{X}_2\}\oplus \mathcal{X}_3\}\semi\{Y(g),Z(h),Q(k),W(m)\}$. Based on these facts  and Theorem 2.1, we can state the following corollary.

\begin{corollary}
\begin{itemize}
\item[(i)] $L_1$ and $L_2$ are isomorphic.
\item[(ii)] The systems \eqref{31MDS} with $a_3=0$,  \eqref{31MDS} with $a_3\neq 0$, $b_1=-2a_1$ and  the system  \eqref{31DS} admit isomoporhic infinite dimensional Lie algebras, expressed as semi-direct sum  of a three-dimensional solvable Lie algebra with a Kac--Moody algebra. 
\item[(iii)] Transformability of \eqref{31MDS} to the equation with $a_3=0$ is possible if $b_1=-2a_1$. This result is also reflected by the fact that in both cases the same Lie algebra is admitted by the equations.
\end{itemize}
\end{corollary}
\begin{remark}
We observe that within the class \eqref{31MDS}, the case $a_3\neq 0$ and $b_1\neq -2a_1$ is particular as the symmetry algebra differentiates and one loses the  symmetry  generated by  $\mathcal{X}_2$ (scaling) or $\mathcal{\tilde X}_2$. 
\end{remark}
As  mentioned in \cite{ozemir2020davey} with reference to  \cite{champagnewinternitz1988},  some  physical symmetries of \eqref{31MDS} can be obtained by restricting $g$, $h$, $k$  and $m$ to be polynomials.  Indeed, $\mathcal{Y}(1)$, $\mathcal{Z}(1)$ and  $\mathcal{Q}(1)$ represent translations along the spatial axes and  $\mathcal{W}(1)$  stands for the phase invariance on the complex function $\psi$. In addition, $\mathcal{Y}(t)$, $\mathcal{Z}(t)$ and  $\mathcal{Q}(t)$ generate one-parameter Lie group of transformations called Galilean boosts in the $x$, $y$, and $z$ directions, respectively.

\section{Exact Solutions}
\subsection{Traveling wave solutions}
We set  $\psi=M(\xi)e^{i\phi(\xi)}$, $w=w(\xi)$, $\xi=e_1 x+e_2 y+e_3 z+t$ in $\eqref{31MDS}$. From \eqref{31MDSb} we obtain
\begin{equation}
w=\frac{b_2(e_1^2+e_3^2)}{e_1^2+b_1 e_2^2+e_3^2}M^2+K_1 \xi+K_0,
\end{equation}
where $K_0$, $K_1$ are arbitrary constants of integration. The real and imaginary parts of \eqref{31MDSa} give 
\begin{subequations}\label{reelsanal}
\begin{eqnarray}
   \label{reel}  &&\gamma \ddot M-\gamma M \dot \phi^2-M\dot \phi-(K_0+K_1\xi)M-\delta M^3=0, \\
   \label{sanal}  && \gamma M \ddot \phi +2\gamma \dot M \dot \phi + \dot M-a_3 M=0
\end{eqnarray}
\end{subequations}
where
\begin{equation}
\gamma = e_1^2+a_1 e_2^2+e_3^2, \qquad \delta = a_2+\frac{b_2(e_1^2+e_3^2)}{e_1^2+b_1 e_2^2+e_3^2}.
\end{equation}
For $\gamma=0$, the solution of \eqref{reelsanal} yields
\begin{equation}
M(\xi)=M_0 e^{a_3\xi}, \quad \phi(\xi)=\phi_0-K_0\xi-\frac{K_1}{2}\xi^2-\frac{\delta M_0^2}{2a_3}e^{2a_3 \xi}
\end{equation}
with arbitrary $M_0$ and $\phi_0$. When $\gamma\neq 0$, multiplying \eqref{sanal} by $M$ and integrating once, we obtain 
\begin{equation}
\dot\phi = \frac{a_3}{\gamma M^2 }\int M^2 d\xi+\frac{C_0}{M^2}-\frac{1}{2\gamma},
\end{equation}
with an  arbitrary constant $C_0$. To be able to proceed with the case $a_3\neq 0$,  let us set $M^2=\dot N$. Then we have 
\begin{equation}
\dot \phi = \frac{a_3}{\gamma}\frac{N}{\dot N}+\frac{C_0}{\dot N}-\frac{1}{2\gamma}
\end{equation}
and \eqref{reel} gives the third order ODE
\begin{equation}\label{eqN}
\dddot N  = \frac{1}{2\dot N}\ddot N ^2+\frac{2\delta}{\gamma}\dot N ^2+\frac{2}{\gamma}(K_1 \xi+K_0-\frac{1}{4\gamma})\dot N + \frac{2}{\dot N} (C_0+\frac{a_3}{\gamma}N)^2.
\end{equation}
When $a_3=0$, setting $\dot N = H$, Eq.  \eqref{eqN} becomes the second order ODE 
\begin{equation}\label{eqH}
\ddot H =\frac{1}{2H}\dot H ^2+\frac{2}{\gamma}(K_1\xi+K_0-\frac{1}{4\gamma})H+\frac{2\delta}{\gamma} H^2+\frac{2C_0^2}{H}.
\end{equation}

For the equations \eqref{eqN} and \eqref{eqH}, we would like to present the following results.
\begin{itemize}[leftmargin=*]
\item[(i)]
When $\delta=0$, multiplying \eqref{eqN} by $\dot N$ and differentiating once  gives the linear equation 
\begin{equation}
N^{(4)}-\frac{4}{\gamma}\Big(K_1\xi+K_0-\frac{1}{4\gamma}\Big)\ddot N-\frac{2K_1}{\gamma}\dot N-\frac{4a_3^2}{\gamma^2}N-\frac{4a_3C_0}{\gamma}=0.
\end{equation}
In this special case, one can obtain the solutions $N$ and hence the functions $\psi$, $w$ in terms of elementary functions.
\item[(ii)] When $\delta \neq 0$, \eqref{eqN} passes the Painlevé test when $a_3=0$. This result was obtained by the package PainleveTestV4-2018.m \cite{heremanwebsite}. 

\item[(iii)] When $K_1=0$, \eqref{eqN} does not include the independent variable hence its order can be reduced by one by the transformation $\dot N =F(N)$, through which one obtains 
\begin{equation}
F''=-\frac{1}{2F} (F')^2+\frac{2\delta}{\gamma}+\frac{2}{\gamma}\big(K_0-\frac{1}{4\gamma}\big)\frac{1}{F}+2(C_0+\frac{a_3}{\gamma}N)^2 \frac{1}{F^3}
\end{equation}
where $F'=\dfrac{dF}{dN}$.
\item[(iv)] In reference \cite{gagnon1989II}, when studying the reductions of the   cubic-quintic nonlinear Schrödinger equation 
\begin{equation}
i\psi_t+\Delta \psi=\tilde a_0+ \tilde a_1 \psi |\psi|^2 +\tilde a_2\psi |\psi|^4
\end{equation}
where $\tilde a_0, \tilde a_1, \tilde a_2 \in \mathbb{R}$, $(\tilde a_1,\tilde a_2)\neq (0,0)$ and $\Delta =\dx^2+\dy^2+\dz^2$, the authors obtain the ODEs
\begin{eqnarray}
\ddot H &=&\frac{1}{2H}\dot H ^2+( a\xi+2\tilde a_0)H+2\tilde a_1 H^2+2\tilde a_2 H^3+\frac{2S_0^2}{H}, \label{eqH1}\\
\ddot H &=&\frac{1}{2H}\dot H ^2+2( \tilde a_0- a)H+2\tilde a_1 H^2+2\tilde a_2 H^3+\frac{2S_0^2}{H}.  \label{eqH2}
\end{eqnarray}

Here $a$ is a group parameter related to the subalgebra used in the reduction and they have the condition $a>0$ for \eqref{eqH1} and $a\in \mathbb{R}$ for \eqref{eqH2}. By a further transformation $H(\xi)=\tilde \lambda(\xi)\tilde H (\tilde\eta)$, $\tilde \eta= \tilde \eta(\xi)$ in \eqref{eqH1} and \eqref{eqH2},  they check whether the new dependent variable $\tilde H (\tilde \eta)$ satisfies one of the second-order canonical equations of Painlevé type classified by Painlevé and Gambier. A full list these equations is available in \cite{ince1956ordinary}.
By a detailed analysis of \eqref{eqH1} and \eqref{eqH2}, they obtain solutions $H(\xi)$ in terms of the second Painlevé transcendent $P_{II}$ for \eqref{eqH1}  when $ \tilde a_2 =0$. For  Eq. \eqref{eqH2}, they obtain solutions in terms of elliptic, hyperbolic and trigonometric functions in case  $ \tilde a_2 =0$. Clearly these solutions also serve for our reduction \eqref{eqH}. However, we do not repeat these results here and suffice to refer to the article \cite{gagnon1989II}.
\end{itemize}

Now, for Eq. \eqref{31MDS} let us look for a solution of the form 
\begin{equation}\label{ansatz}
	\psi ={\Omega}(\eta) e^{i(\theta_1 x+\theta_2 y+\theta_3 z+ t)}, \quad w= \chi (\eta), \quad \eta=\eta_1 x+\eta_2y+\eta_3 z+ t
\end{equation}
where $\eta_i$, $\theta_i$, $i=1,2,3$   are real constants.
We obtain 
\begin{align}
 \left(\eta_1 ^2+a_1 \eta_2 ^2+\eta_3^2  \right)\Omega''  - \left(1+\theta _1^2+a_1 \theta _2^2+\theta _3^2+\chi\right)\Omega-a_2\Omega^3 
	 &=0, \label{red1}\\ 
	(\eta_1^2+b_1 \eta_2^2+\eta_3^2) \chi''- b_2 (\eta_1^2+\eta_3^2) (\Omega^2)''&=0,  \label{red2}\\
	(1 + 2 \eta_1 \theta_1 + 2 a_1 \eta_2 \theta_2 + 
  2 \eta_3 \theta_3) \Omega' -a_3 \Omega &=0.  \label{red3}
\end{align}
\textbf{Case I.}  When $a_3\neq 0 $, \eqref{red3} requires
\begin{equation}\label{solP}
	\Omega=c_0  e^{c_1 \eta}, \qquad c_1= \frac{a_3}{1 + 2 \eta_1 \theta_1 + 2 a_1 \eta_2 \theta_2 + 
  2 \eta_3 \theta_3}.
\end{equation}
We integrate \eqref{red2} twice by taking the constant of first  integration as zero and  get
\begin{equation}\label{eqfi2}
	\chi=d_1\Omega^2+d_0, \qquad d_1=\frac{b_2 (\eta_1^2+\eta_3^2)}{\eta_1^2+b_1 \eta_2^2+\eta_3^2}
\end{equation}
where $d_0$ is the integration constant.  Finally, equation \eqref{red1} determines the constants
\begin{equation}
    d_0=\frac{a_3^2 \left(\eta_1 ^2+a_1 \eta_2 ^2+\eta_3 ^2\right)}{(1 + 2 \eta_1 \theta_1 + 2 a_1 \eta_2 \theta_2 + 
  2 \eta_3 \theta_3)^2}-\left(1+\theta _1^2+a_1 \theta _2^2+\theta _3^2\right), \qquad a_2=-\frac{b_2(\eta_1^2+\eta_3^2)}{\eta_1^2+b_1 \eta_2 ^2+\eta_3 ^2}.
\end{equation}
Therefore, the solution is 
\begin{align}
\psi(x,y,z,t)&=c_0 e^{c_1 (\eta_1 x+\eta_2y+\eta_3 z+ t)} e^{i(\theta_1 x+\theta_2 y+\theta_3 z+ t)},\\
w(x,y,z,t)&=d_0+d_1 c_0^2  e^{2c_1 (\eta_1 x+\eta_2y+\eta_3 z+ t)}.
\end{align}
\textbf{Case II.} When $a_3=0$, we have the set of equations 
\begin{align}
 \left(\eta_1 ^2+a_1 \eta_2 ^2+\eta_3^2  \right)\Omega''  - \left(1+\theta _1^2+a_1 \theta _2^2+\theta _3^2+\chi\right)\Omega-a_2\Omega^3 
	 &=0, \label{red12}\\ 
	(\eta_1^2+b_1 \eta_2^2+\eta_3^2) \chi''- b_2 (\eta_1^2+\eta_3^2) (\Omega^2)''&=0  \label{red22}
\end{align}
with the condition
\begin{align}
1 + 2 \eta_1 \theta_1 + 2 a_1 \eta_2 \theta_2 + 
  2 \eta_3 \theta_3 =0
\end{align} 
so that $\Omega'\neq 0$.   
Integration of \eqref{red22} gives
\begin{equation}
\chi(\eta) =  d_1 \Omega^2(\eta) + \chi_0
\end{equation}
where $\chi_0$ is arbitrary and $d_1$ is as before.
Substituting 	$\chi$  into \eqref{red12} we obtain
\begin{equation}\label{power22}
	(\eta _1^2+a_1 \eta _2^2+\eta _3^2) \Omega '' =( a_2+d_1) \Omega ^3+(1+\chi_0+\theta _1^2+a _1 \theta _2^2+\theta _3^2)\Omega
\end{equation}
and, in a compact form,
\begin{equation}\label{power1}
	 \Omega '' =A \Omega ^3+L\Omega
\end{equation}
where 
\begin{equation}\label{powerex}
	 A=\frac{a_2+d_1}{\eta _1^2+a_1 \eta _2^2+\eta _3^2}, \qquad L=\frac{1+\chi_0+\theta _1^2+a _1 \theta _2^2+\theta _3^2}{\eta _1^2+a_1 \eta _2^2+\eta _3^2}. 
\end{equation}
 \noindent Multiplying both sides of the Eq. \eqref{power1} by $\Omega' $, we obtain
\begin{equation}\label{power}
	(\Omega')^2=J \Omega ^4+L\Omega^2+K
\end{equation}
where $J=A/2$. Integration of Eq. \eqref{power} can be performed by elementary operations. For completeness, we mention results from \cite{gonul2022benney}. In what follows we assume $J\neq 0$.

\noindent \textbf{(A)} When $K=0$ and $L\neq 0$ \eqref{power} reduces to the equation, 
	\begin{equation}
		\frac{d \Omega}{\Omega \sqrt{J \Omega^2+L}} = \varepsilon d \eta,
	\end{equation}
	  $\varepsilon =\mp 1$. This integral can be evaluated in three different cases. 
   
\noindent   \textbf{(A.1)} If $J < 0$ and $L>0$,   we integrate and find 
	\begin{align}
		\label{sech}\Omega(\eta) &=\sqrt{-\frac{L}{J}}\ \mathrm{sech}\Big(\sqrt{L}(\eta+\eta_0)\Big),  \\
		\label{sech2}\mathrm{\chi}(\eta) &=-\frac{d_1L}{J}\mathrm{sech}^2 \Big(\sqrt{L}(\eta+\eta_0)\Big)+\chi_0.
	\end{align}
 For $\chi_0=0$,  $ \Omega(\eta)$ and  $\chi(\eta)$ decay to zero at infinity except specific directions. The graphs  of  $\Omega(\eta)$ and $\chi(\eta)$ on any plane in space  are line solitons.  
 \begin{figure}[h]
	\centering
	\begin{subfigure}[]
		\centering
		\includegraphics[scale=0.5]{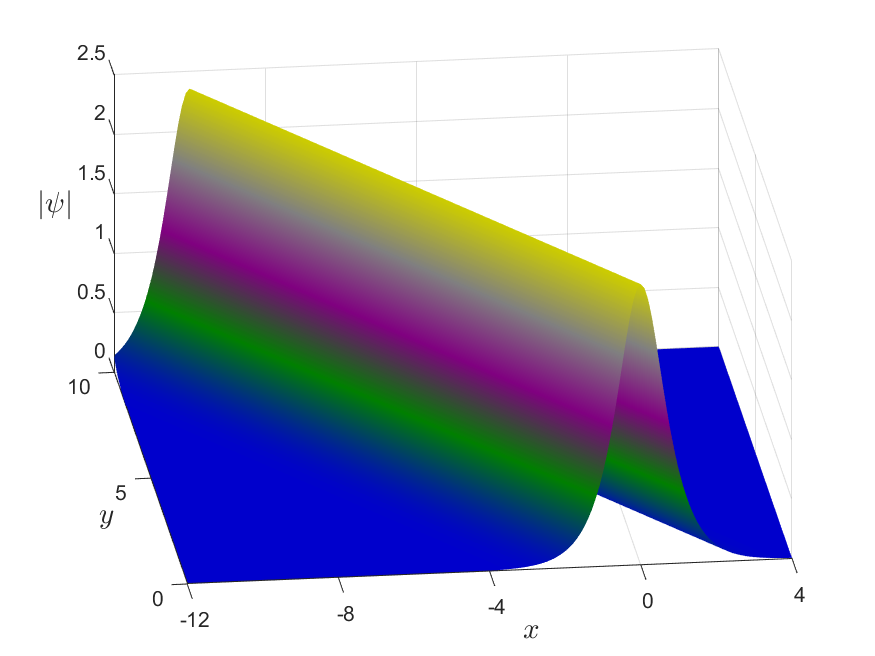}
		\label{fig:comparisondiagram}
	\end{subfigure}
	\begin{subfigure}[]
		\centering
		\includegraphics[scale=0.5]{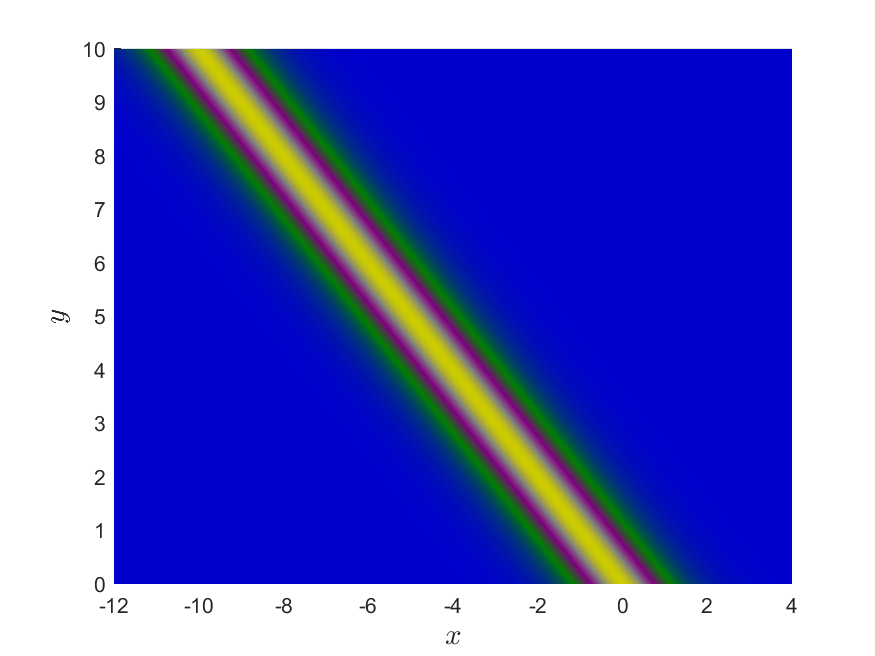}
		\label{fig:comparisondiagram}
	\end{subfigure}
\begin{subfigure}[]
	\centering
	\includegraphics[scale=0.5]{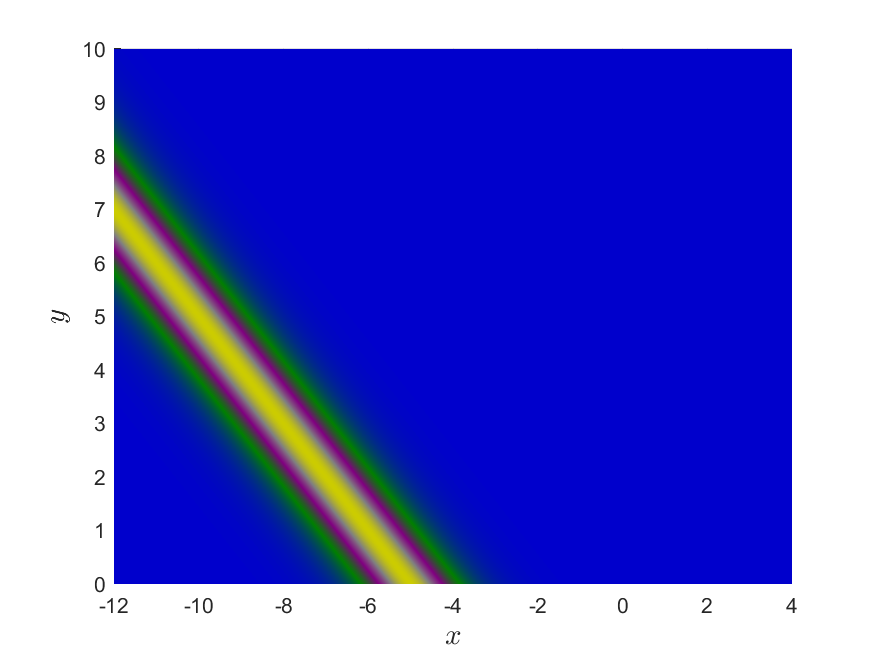}
	\label{fig:comparisondiagram}
\end{subfigure}
\begin{subfigure}[]
	\centering
	\includegraphics[scale=0.5]{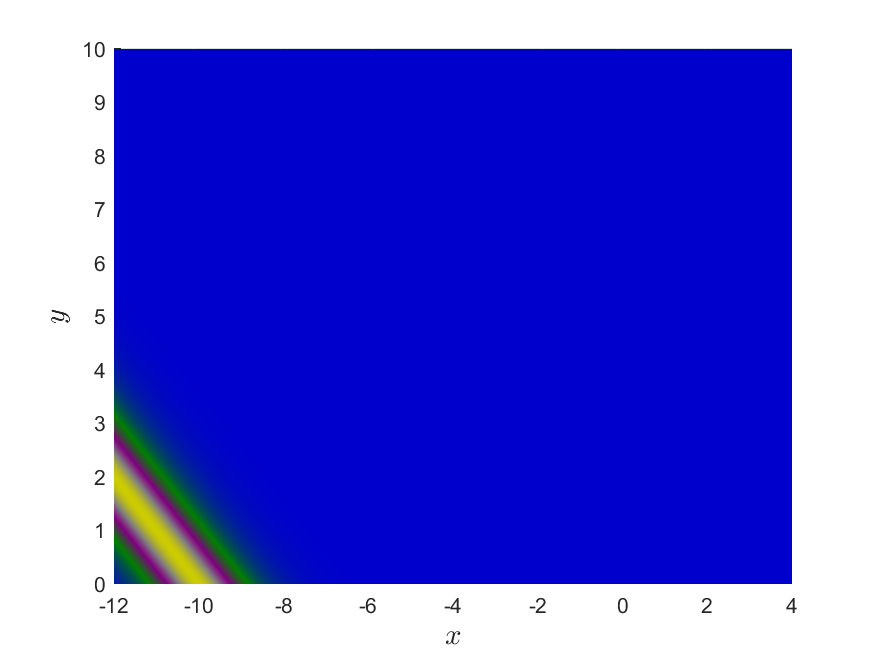}
	\label{fig:comparisondiagram}
\end{subfigure}
\caption{Solution \eqref{sechn} on the plane $z=0$; (a) $t=0$, (b) $t=0$, (c) $t=5$, (d) $t=10$. }
\end{figure}
\begin{figure}[h]
	\centering
	\begin{subfigure}[]
		\centering
		\includegraphics[scale=0.5]{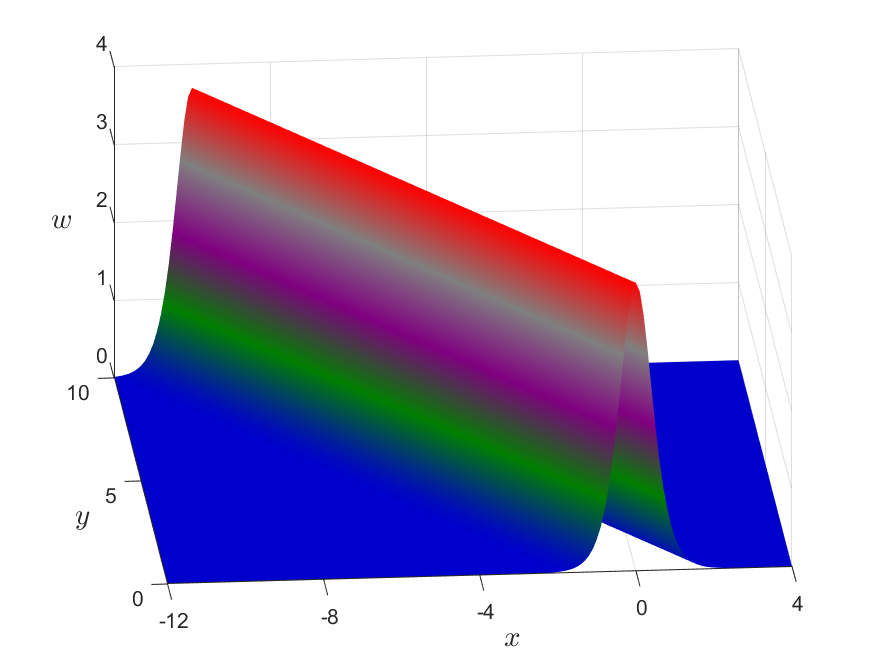}
	\end{subfigure}
	\begin{subfigure}[]
		\centering
		\includegraphics[scale=0.5]{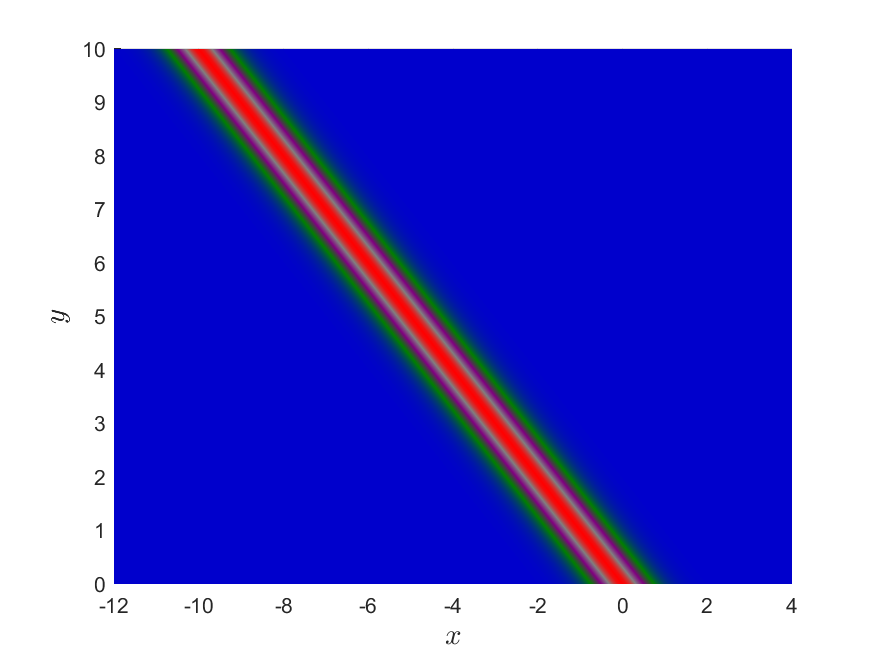}
		\label{fig:comparisondiagram}
	\end{subfigure}
\begin{subfigure}[]
	\centering
	\includegraphics[scale=0.5]{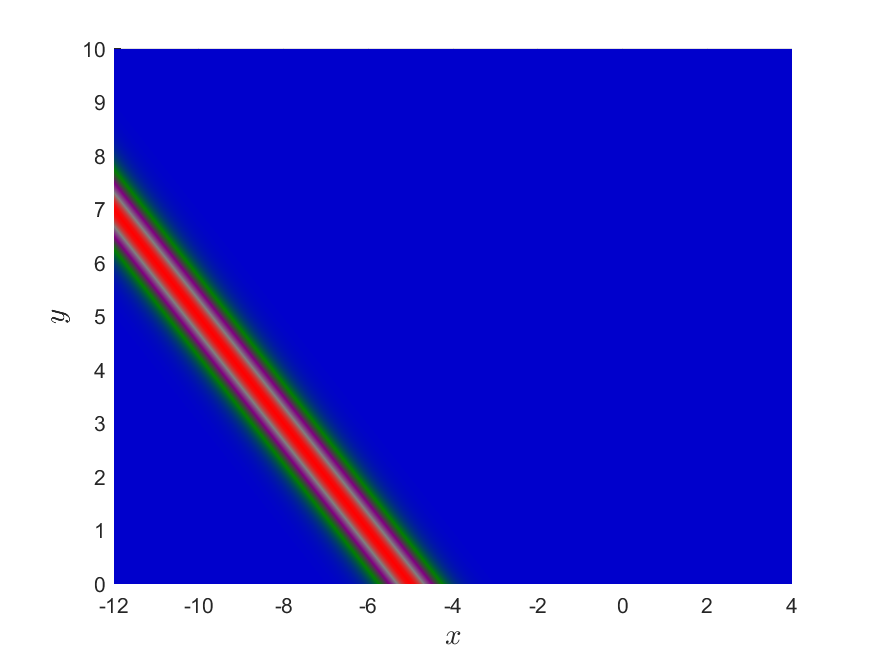}
	\label{fig:comparisondiagram}
\end{subfigure}
\begin{subfigure}[]
	\centering
	\includegraphics[scale=0.5]{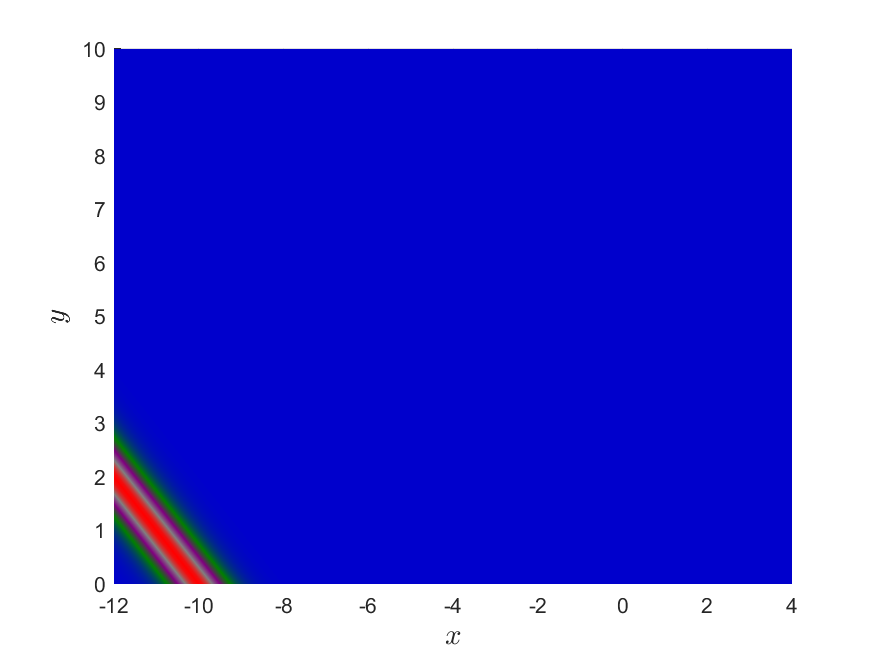}
	\label{fig:comparisondiagram}
\end{subfigure}
\caption{Solution \eqref{sech2n} on the plane $z=0$; (a) $t=0$, (b) $t=0$, (c) $t=5$, (d) $t=10$. }
\end{figure}

\noindent\textbf{ (A.2)}  In case $J>0$ and $L>0$, we obtain
	\begin{align}
		\Omega(\eta)&=-\varepsilon\sqrt{\frac{L}{J}}\ \mathrm{cosech}\Big(\sqrt{L}(\eta+\eta_0)\Big),\\
		\chi(\eta)&=\frac{d_1L}{J}\mathrm{cosech}^2 \Big(\sqrt{L}(\eta+\eta_0)\Big)+\chi_0.
	\end{align}
\vspace{.3cm}

\noindent\textbf{ (A.3)}  When $J>0$ and $L<0$, we get
	\begin{align}
		\Omega(\eta)&=\sqrt{-\frac{L}{J}}\ \mathrm{sec}\Big(\sqrt{-L}(\eta+\eta_0)\Big),\\
		\chi(\eta)&=-\frac{d_1L}{J}\mathrm{sec}^2 \Big(\sqrt{-L}(\eta+\eta_0)\Big)+\chi_0.
	\end{align}
\vspace{.3cm}

\noindent {\textbf{(B)}  When $ \Delta=L^2-4 JK=0$, Eq.   \eqref{power} turns into
		\begin{equation}
			\frac{d\Omega}{2J\Omega^2+L}=\frac{\varepsilon}{2\sqrt{J}}\ d\eta
		\end{equation}
with  		 $\varepsilon =\mp 1$.

\noindent\textbf{ (B.1)}  If $L<0$ we obtain,
		\begin{align}
			&\Omega(\eta)=-\varepsilon\sqrt{\frac{-L}{2J}}\tanh \Big[\sqrt{-\frac{L}{2}} (\eta+\eta_0)\Big], \label{tanjant1}\\
			&\mathrm{\chi}(\eta)=\frac{d_1L}{2J}\mathrm{sech}^2 \Big[\sqrt{-\frac{L}{2}} (\eta+\eta_0)\Big]-\frac{d_1L}{2J}+\chi_0.    \label{tanjant2}
		\end{align}
	
 For $\chi_0=d_1 L /(2J)$,   $\chi(\eta)$ becomes  zero at infinity except special directions. On any plane in space, the graph of  $\Omega(\eta)$ is a kink-soliton and the graph of $\chi(\eta)$ is a line-soliton.

\noindent\textbf{(B.2)}  If $L>0$ we obtain,
		\begin{align}
		&\Omega(\eta)=\varepsilon\sqrt{\frac{L}{2J}}\tan\Big( \sqrt{\frac{L}{2}}(\eta+\eta_0)\Big),\\	
		&\mathrm{\chi}(\eta)=\frac{ d_1 L}{2J}\tan^2\Big(\sqrt{\frac{L}{2}}(\eta+\eta_0)\Big)+\chi_0.
         \end{align}
 When $\Delta=L^2-4 JK\neq 0$, exact solutions of elliptic type or periodic functions are obtained through integration of \eqref{power}. \cite{gonul2022benney} contains results in this regard.

 In this family of analytical solutions to the (3+1)-dimensional DS system, we would like to illustrate the couple $(\Omega,\chi)$ that is presented in \eqref{sech} and \eqref{sech2}. The associated constants are $a_1=b_1=b_2=\eta_1=\eta_2=\eta_3=\theta_1=\theta_2=1$, $a_2=-4$, $d_1=2/3$, $\theta_3=-5/2$ and  $\eta_0=\chi_0=0$, which results in 
 \begin{align}
\label{sechn}
{\hat\Omega}(\eta) &=\sqrt{\frac{111}{20}}\ \mathrm{sech}\Big(\sqrt{\frac{37}{12}}(x+y+z+t)\Big),   \\
\label{sech2n}
\hat \chi(\eta) &=\frac{37}{10} \,  \, \mathrm{sech}^2\Big(\sqrt{\frac{37}{12}}(x+y+z+t)\Big).
\end{align}\vspace{0.2cm}
 Let us note that with this set of parameters, the solution to   \eqref{31MDS} with $a_3=0$ will be 
 \begin{equation}
\psi(x,y,z,t)=\hat\Omega(\eta)e^{i(x+y-\frac{5}{2} z+ t)}, \qquad  w(x,y,z,t)=\hat \chi(\eta).
 \end{equation}
 We plot $|\psi|=\hat \Omega(\eta)$ in Figure 1 and $w=\hat \chi(\eta)$ in Figure 2 on the plane $z=0$ for the times $t=0,5,10$  illustrating the line solitons couple both  propagating on the $xy$-plane in the direction of the  vector $(-1,-1)$ with a speed of $1$ but with different amplitudes. 
 
\subsection{Stability Analysis}
\begin{figure}[h!]
	\centering
	\begin{subfigure}[]
		\centering
		\includegraphics[scale=0.50]{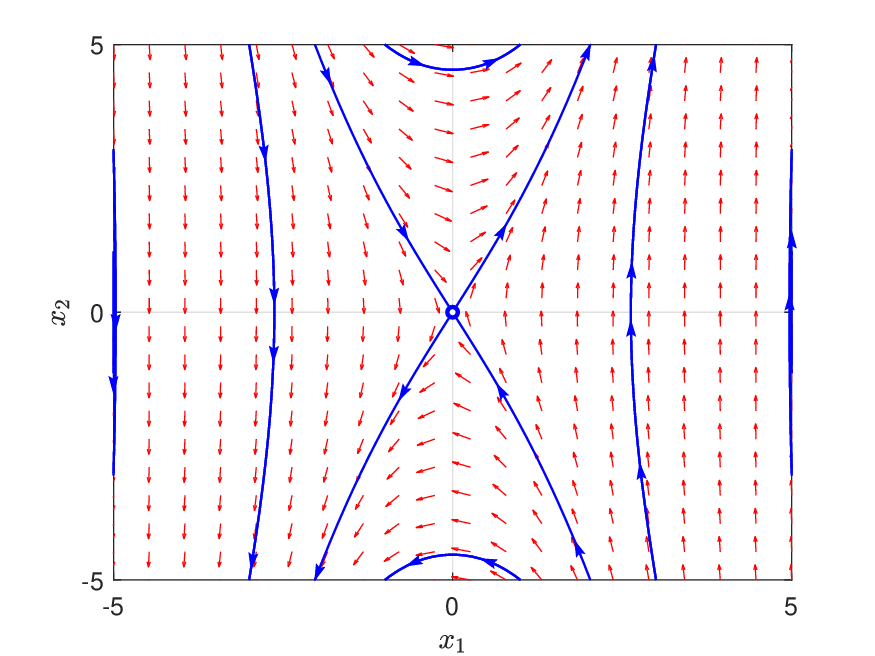}
	\end{subfigure}
 \begin{subfigure}[]
	\centering
	\includegraphics[scale=0.50]{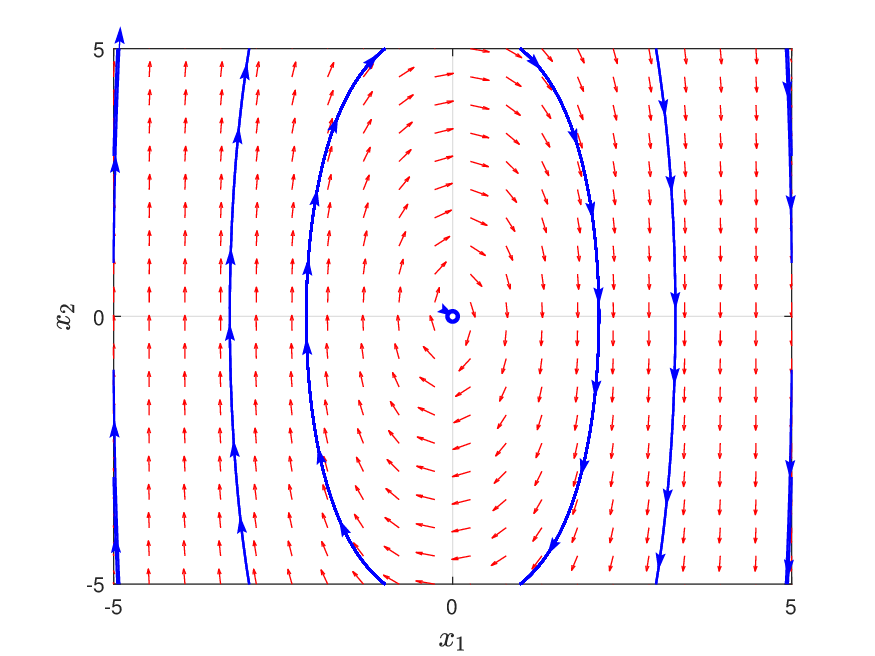}
	\label{fig:comparisondiagram}
\end{subfigure}
	\begin{subfigure}[]
		\centering
		\includegraphics[scale=0.50]{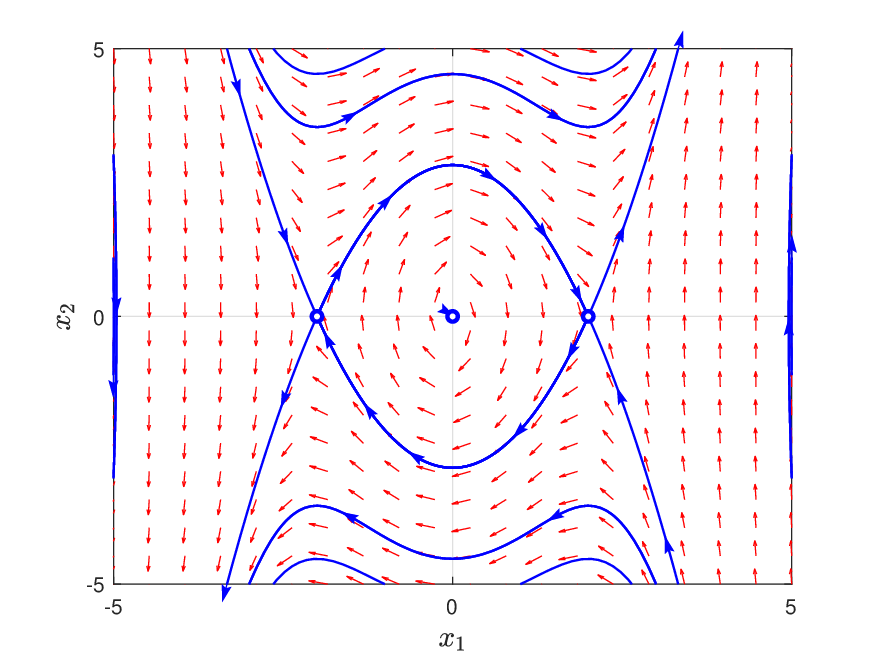}
		\label{fig:comparisondiagram}
	\end{subfigure}
\begin{subfigure}[]
	\centering
	\includegraphics[scale=0.50]{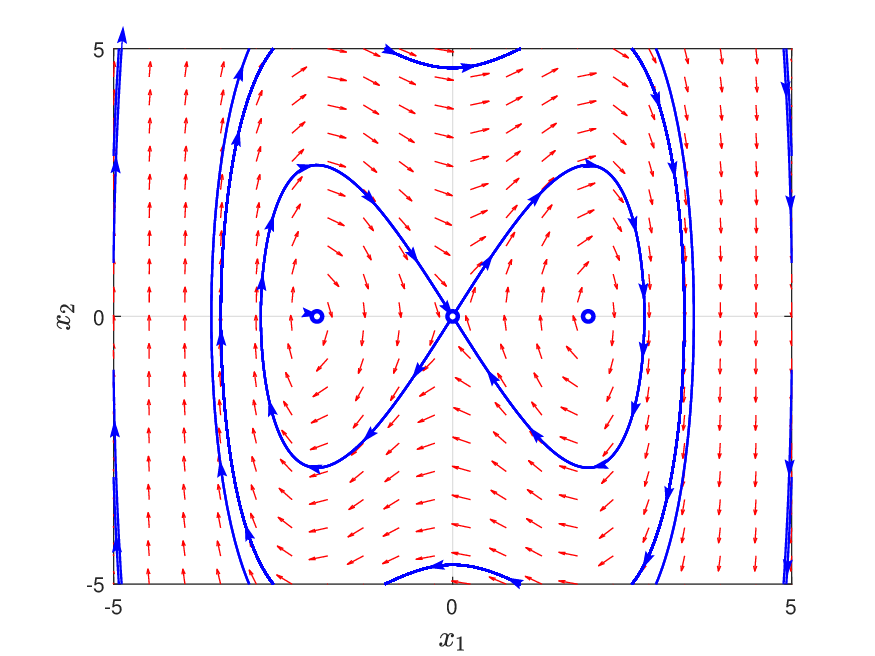}
	\label{fig:comparisondiagram}
\end{subfigure}
\begin{subfigure}[]
		\centering
		\includegraphics[scale=0.50]{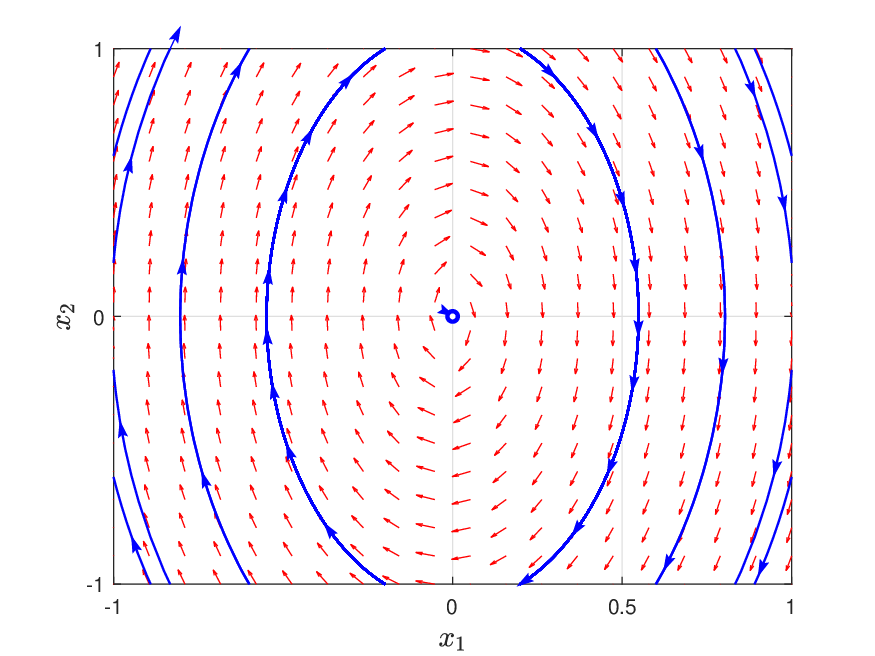}
		\label{fig:comparisondiagram}
	\end{subfigure}
\begin{subfigure}[]
	\centering
	\includegraphics[scale=0.50]{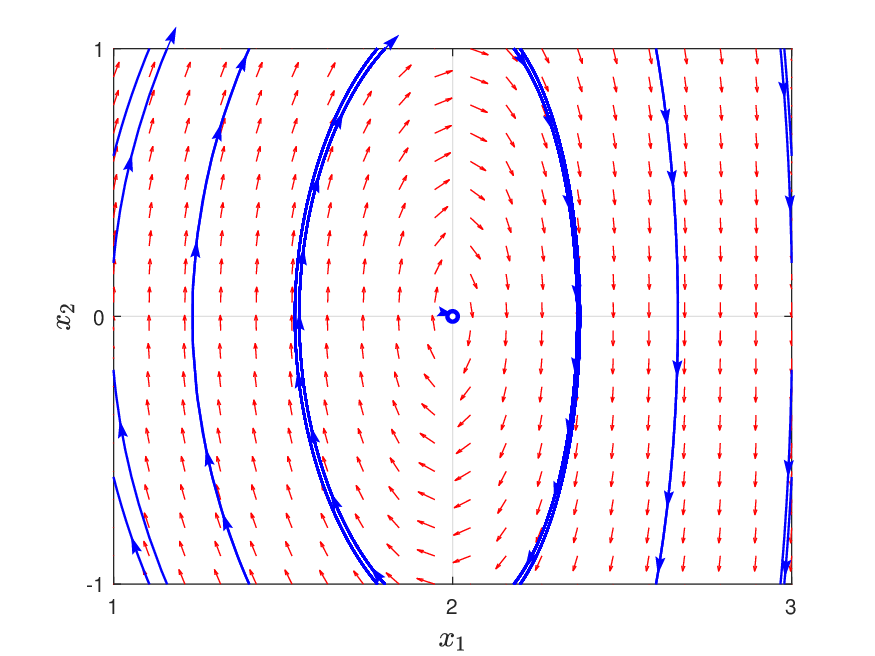}
	\label{fig:comparisondiagram}
\end{subfigure}

\caption{ (a) $A=1$, $L=4$ (b) $A=-1$, $L=-4$, (c) and (e) $A=1$, $L=-4$, (d) and (f) $A=-1$, $L=4$.  }
\end{figure}

In the previous subsection, we found some exact solutions to the equation   \eqref{power1}, which is the ODE satisfied by solutions of the form \eqref{ansatz} to the PDE \eqref{31MDS} with $a_3=0$.  Now, we would like to perform a stability analysis for the solutions to \eqref{power1} by reducing it to a system of first order ODEs by introducing the variables $x_1=\Omega$, $x_2=\Omega'$, which results in 
\begin{subequations}\label{8D}
\begin{eqnarray}
\frac{dx_1}{d\eta}&=&x_2,\\
\frac{dx_2}{d\eta}&=&Ax_1^3+Lx_1.
\end{eqnarray}
\end{subequations}
We see in \cite{selima2018applicable} that  this system of equations arises for the stability analysis of exact solutions to a DS system in $(3+1)$ dimensions similar to \eqref{31MDS} with $a_3=0$, the case we are analysing here and below we provide a detailed discussion. 
As integration of \eqref{power1} is trivial when $L=0$, below we assume $L\neq 0$. 
On the $x_1x_2$-plane, let us define the points
\begin{equation}
E_1{(0,0)}, \quad E_2{(-\sqrt{-\dfrac{L}{A}},0)}, \quad E_3{(\sqrt{-\dfrac{L}{A}},0)}.  
\end{equation}
The equilibrium points of this nonlinear system realizes in two different cases. If \, $\dfrac{L}{A}>0$, there is only one equilibrium point $E_1$. If \, $\dfrac{L}{A}<0$, there are three equilibrium points $E_1$, $E_2$, $E_3$.  
The Jacobian of the system is 
\begin{equation} 
J=\begin{pmatrix}
0 & 1 \\
3Ax_1^2+L & 0 
\end{pmatrix}.
\end{equation} 
At the equilibrium point $E_1$, we have the characteristic equation 
\begin{equation}
\Big|J|_{E_1}-\lambda I\Big|=\lambda^2-L=0.
\end{equation}
Therefore, if $L>0$, we have the eigenvalues $\lambda_{1,2}=\mp \sqrt{L}$ and $E_1$ is a saddle point. If $L<0$, eigenvalues are obtained to be $\lambda_{1,2}=\mp \sqrt{-L}i$ and linear stability analysis does not conclude a result. 

At the equilibrium points $E_2$ and $E_3$, eigenvalues of $J$ satisfy $\lambda^2= -2L$. So, when $L<0$ and $A>0$, the eigenvalues are $\lambda_{1,2}=\mp  \sqrt{-2L}$ and the equilibrium points are saddle points. 
If $L>0$ and $A<0$, the eigenvalues are found as $\lambda_{1,2}=\mp  \sqrt{2L}i$ and linear stability analysis does not yield any result. We keep the stability analysis in the case of pure imaginary eigenvalues out of the scope of this article and suffice to draw the phase diagrams only.  

We consider specific values of $A$ and $L$, which we see can be assumed for specific values of the constants appearing in the formulae available in \eqref{powerex}.  In Figure 3(a) with $A=1$ and $L=4$, there is the unique equilibrium  $E_1(0,0)$ which is a saddle point. In Figure 3(b) where $A=-1$ and $L=-4$, again there is the unique equilibrium $E_1(0,0)$ and we see that it appears as a centre point.   In case $A=1$, $L=-4$ and $A=-1$, $L=4$, plotted in Figure 3(c) and 3(d), respectively,  there are three equilibrium points $E_1(0,0)$, $E_2(-2,0)$ and $E_3(2,0)$.  Figure 3(e) zooms at $E_1$ of Figure 3(c) and Figure 3(f) focuses on $E_3$ of Figure (3d). Therefore, in Figures 3(c,e) we see that $E_1$ is a centre whereas $E_2$ and $E_3$ are saddles. Figures 3(d,f) illustrate $E_1$ as a saddle and $E_2$, $E_3$ as centre points.  Phase portraits in Figure 3 were obtained by using  the package available at \cite{yuzhang}.

\section{Conclusion}

In this work we analysed a modified Davey--Stewartson sytem which was reported in the literature as the governing equations of nonlinear waves in a plasma medium in three space dimensions. This system includes an additional term, which puts it in a distinct place amongst other DS-type systems of PDEs. We wondered whether this term can be removed by a point transformation of the dependent and the independent variables. We showed that, if a specific condition is satisfied for the parameters of the system, the answer is affirmative.  After that, for both of the cases whether this removal can be achieved or not, we investigated the Lie symmetry algebra of the modified DS system and presented the structure of the invariance algebra, mentioning connections with the available literature. We also attempted to obtain some exact solutions in various cases. In special, for the wave components virating in orthogonal directions we obtained a line soliton couple and a kink soliton-line soliton couple.

The system we considered was derived based on concrete physical grounds, to the best of our knowledge, and  has not been considered in the literature from the standpoint we have been in this manuscript. We believe the results we presented will be useful to the community who work on these classes of PDEs and stimulate further research in other directions.

\section*{Acknowledgements}
This work was supported by Scientific Research Projects Department of Istanbul Technical University, Project Number: THD-2024-45388.  This article was in initial stages when C. Ozemir was visiting University of Granada with this support; warm hospitality  by Prof. Pedro Torres and the Mathematics Department is greatly acknowledged.

%%%\bibliographystyle{unsrt}
%%%%\bibliographystyle{plain}

%%%%%\bibliography{RefsMDS31}

\end{document}